\documentclass[aps,prl,superscriptaddress,amsmath,amssymb,twocolumn,fleqn,nofootinbib]{revtex4}

\usepackage[utf8]{inputenc}
\usepackage{graphicx}
\usepackage{amsmath}
\usepackage{amssymb}
\usepackage{amsfonts}
\usepackage{lscape}
\usepackage{hyperref}
\usepackage{url}
\usepackage{epsfig}
\usepackage{color}
\usepackage{bm}
\usepackage{tabularx}
\usepackage{braket}
\usepackage{comment}
\newcommand{\be}{\begin{equation}}
\newcommand{\ee}{\end{equation}}
\newcommand{\ba}{\begin{eqnarray}}
\newcommand{\ea}{\end{eqnarray}}
\newcommand{\nn}{\nonumber}

\newcommand{\Msun}{M_\odot}
\newcommand{\OGW}{\Omega_{\rm GW}}
\newcommand{\ODM}{\Omega_{\rm DM}}
\newcommand{\OM}{\Omega_{\rm M}}

\newcommand{\fPBH}{f_{\rm PBH}}
\newcommand{\dloc}{\delta_{\rm loc}}

\begin{document}

\thispagestyle{empty}

\title{The stochastic gravitational wave background from close hyperbolic encounters of primordial black holes in dense clusters}

\author{Juan Garc\'ia-Bellido}\email[]{juan.garciabellido@uam.es}
\author{Santiago Jaraba}\email[]{santiago.jaraba@uam.es}

\affiliation{Instituto de F\'isica Te\'orica UAM-CSIC, Universidad Auton\'oma de Madrid, Cantoblanco 28049 Madrid, Spain}

\author{Sachiko Kuroyanagi}\email[]{sachiko.kuroyanagi@csic.es}
\affiliation{Instituto de F\'isica Te\'orica UAM-CSIC, Universidad Auton\'oma de Madrid, Cantoblanco 28049 Madrid, Spain}
\affiliation{Department of Physics and Astrophysics, Nagoya University, Nagoya, 464-8602, Japan}

\date{\today}

\begin{abstract}
The inner part of dense clusters of primordial black holes is an active environment where multiple scattering processes take place. Some of them give rise from time to time to bounded pairs, and the rest ends up with a single scattering event. The former eventually evolves to a binary black hole (BBH) emitting periodic gravitational waves (GWs), while the latter with a short distance, called close hyperbolic encounters (CHE), emits a strong GW burst. We make the first calculation of the stochastic GW background originating from unresolved CHE sources. Unlike the case for BBH, the low-frequency tail of the SGWB from CHE is sensitive to the redshift dependence of the event rate, which could help distinguish the astrophysical from the primordial black hole contributions. We find that there is a chance that CHE can be tested by third-generation ground-based GW detectors such as Einstein Telescope and Cosmic Explorer.

\end{abstract}
\maketitle

\section{I. Introduction}

Primordial black holes (PBHs) may have formed in the early Universe ~\cite{Zeldovich:1967lct,Hawking:1971ei,Carr:1974nx,Carr:1975qj,Chapline:1975ojl}. They can be formed by the gravitational collapse of overdense regions, caused by high peaks in the primordial curvature power spectrum generated during inflation~\cite{Dolgov:1992pu,Carr:1993aq,Carr:1994ar,Ivanov:1994pa,GarciaBellido:1996qt}. There are also other mechanisms to produce PBHs through e.g., phase transitions~\cite{Jedamzik:1996mr}, scalar field instabilities~\cite{Khlopov:1985jw}, the collapse of cosmic strings~\cite{Polnarev:1988dh}, etc. PBHs have been studied for decades as they may account for all or part of the dark matter (DM) in the Universe.

There is not yet definite proof of the existence of PBHs, but recently, gravitational wave (GW) observation of binary black hole (BBH) mergers is providing rich information on the BH population~\cite{LIGOScientific:2020ibl}. Some analysis based on the mass and rate distributions~\cite{Carr:2019kxo,Jedamzik:2020ypm,Jedamzik:2020omx,Clesse:2020ghq,Hall:2020daa,Hutsi:2020sol,Franciolini:2021tla} or spin properties~\cite{Fernandez:2019kyb,Garcia-Bellido:2020pwq} suggests that the observed BBHs could be primordial origin. 

Another approach to probe PHBs is to search a stochastic GW background (SGWB), which can be formed both at the PBH formation~\cite{Saito:2008jc,Bugaev:2009zh,Saito:2009jt} and by the superposition of GWs from BBHs~\cite{Mandic:2016lcn,Clesse:2016ajp,Raidal:2017mfl,Chen:2018rzo,Wang:2019kaf}. The LIGO and Virgo detectors have been improving the upper limit on the amplitude of SGWB~\cite{LIGOScientific:2021yuo}, and constraints on PBHs through SGWB have been discussed~\cite{Wang:2016ana,Kapadia:2020pnr,Mukherjee:2021ags}. In the future, the upgraded version of the LIGO-Virgo-KAGRA detector network~\cite{LVK:2013pob} (and later with LIGO-India) and next-generation GW experiments such as Einstein Telescope (ET)~\cite{Punturo:2010zz}, Cosmic Explorer (CE)~\cite{Reitze:2019iox}, LISA~\cite{LISA:2017pwj}, TianQin~\cite{TianQin:2015yph}, Taiji~\cite{Hu:2017mde}, DECIGO~\cite{Kawamura:2020pcg}, and so on will allow us to search SGWBs with greater sensitivities for a wide range of frequencies. 

In this paper, we propose an important additional source of a SGWB, which is formed by overlapped GW bursts from close hyperbolic encounters (CHEs). When considering two interacting BHs, it is possible that BHs will not end up in bound systems depending on the initial condition but instead produce single scattering events. GW bursts from such unbounded interacting systems can be observed by GW experiments and have been studied in the literature~\cite{Kocsis:2006hq,OLeary:2008myb,Capozziello:2008ra,DeVittori:2012da,Garcia-Bellido:2017knh,Garcia-Bellido:2017qal,Grobner:2020fnb,Mukherjee:2020hnm}. In fact, the dense environment at the center of the clusters can enhance the rate of events with an eccentricity near to unity~\cite{Trashorras:2020mwn}, leading to a strong GW burst. Furthermore, if the interaction is strong enough, they can produce interesting dynamics, such as spin induction~\cite{Jaraba:2021ces, Nelson:2019czq} and subsequent mergers~\cite{Healy:2009zm}, etc. If a sufficient number of events occur in the Universe, they overlap and form a SGWB. We make the first estimation of the SGWB amplitude from CHEs and discuss its detectability in future GW experiments, comparing it with the one from BBHs.

\section{II. Stochastic background of GW}
First, we briefly provide the formulations for calculating the SGWB spectra. The amplitude of a SGWB is commonly characterized by $\Omega_{\rm GW}\equiv (d\rho_{\rm GW}/d\ln f)/\rho_c$, where $f$ is the frequency in the observer frame, $\rho_{\rm GW}$ is the energy density of GWs, and $\rho_c\equiv 3H_0^2/(8\pi G)$ is the critical density of the Universe with $G$ being the Newton constant. We use the Hubble constant normalized by $H_0=70~h_{70}~{\rm km/s/Mpc}$. For a SGWB of point source origin, the GW spectrum is commonly computed by
\be
\label{eq:OGW}
\OGW(f) = \frac{1}{\rho_c}\int_{0}^\infty \!dz\,\frac{N(z)}{1+z}\,\frac{dE_{\rm GW}}{d\ln f_r}\,,
\ee
where $f_r = (1+z)f$ is the GW frequency in the source frame and $dE_{\rm GW}/d\ln f_r$ is the GW energy emission per logarithmic frequency bin in the source frame. Here, $N(z)$ is the number density of GW events at redshift $z$ and given by
\be
N(z) = \frac{\tau(z)}{(1+z)H(z)} \,,
\ee
where $H(z)$ is the Hubble expansion rate and $\tau(z)$ is the merger rate per unit time per comoving volume, in units of ${\rm yr}^{-1}{\rm Gpc}^{-3}$. The merging rate depends on the two progenitor masses $m_1$ and $m_2$, and for a broad mass distribution of BHs, $\tau(z)$ should be replaced by
\be
\tau(z) = \int\!\!\int \frac{d m_1}{m_1}\,\frac{d m_2}{m_2}\,
\frac{d\tau}{d\ln m_1\,d\ln m_2} \,,
\ee
where $d\tau/(d\ln m_1 d \ln m_2)$ is now the merging rate per logarithmic mass interval.

\subsection{II.1 Binary BHs}
There are two different channels for PBH binary formation. Early formation of gravitationally bound objects due to the tidal torques from other PBHs takes place before the matter-radiation  equality~\cite{Nakamura:1997sm,Sasaki:2016jop}, while in the late Universe, where the clustering of PBHs becomes important, PBH binaries can be created by dynamical capture~\cite{Quinlan:1989,Mouri:2002mc}. It is still unclear which gives the dominant contribution as it depends on the mass distribution of PBHs and the clustering nature of PBHs~\cite{Clesse:2016ajp}. Here, however, we describe the BBH formation through the latter mechanism since it is closely related to CHE. 

Let us consider an interaction of two BHs where the masses are $m_1$ and $m_2$, their relative velocity at infinity is $v_0$, and the impact parameter $b$ and the eccentricity $e$ are related by $b=(GM/v_0^2)\,\sqrt{e^2-1}$ where $M=m_1+m_2$. The condition that a binary BH pair forms by a close encounter is that the energy loss due to GW emission exceeds the kinetic energy $E_\infty =1/2 \,\mu v_0^2$ where $\mu\equiv m_1 m_2/(m_1+m_2)$ is the reduced mass. The energy loss by the GW emission is given by
\ba
\Delta E &=& -\frac{8}{15}\frac{G^{7/2}}{c^5}\frac{M^{1/2}m_1^2m_2^2}{r_{\rm min}^{7/2}}f(e)\,,\\
f(e) &\equiv& \frac{1}{(1+e)^{7/2}}\left[24\cos^{-1}(-1/e)
\left(1+\frac{73}{24}e^2+\frac{37}{96}e^4\right)\right. \nn \\[1mm]
&&+\left.\sqrt{e^2-1}\left(\frac{301}{6}+
\frac{673}{12}e^2\right)\right]\,,
\ea
where $r_{\rm min}$ is the distance of closest approach, which can be expressed in terms of $v_0$ as $r_{\rm min}=(GM/v_0^2)(e-1)$. Consequently, the condition $E_\infty\leq |\Delta E|$ gives~\cite{Quinlan:1989,Mouri:2002mc}
\ba
G(e) &\geq& \frac{15}{16} \frac{M^2}{m_1\,m_2}\frac{c^5}{v_0^5}\,,  
\label{eq:condition}
\\[1mm]
G(e) &\equiv& \frac{f(e)}{(e-1)^{7/2}}\nn \\[2mm]
 &=& \left\{
    \begin{array}{lr}
    {\displaystyle
     \frac{425\pi}{4}\,(e^2-1)^{-7/2} } & {\rm for} \ e\approx 1 \,, \\[3mm]
     {\displaystyle
     \frac{37\pi}{8}\,(e^2-1)^{-3/2} } & {\rm for} \ e\gg 1 \,.
    \end{array}
    \right.
\ea

When Eq.~\eqref{eq:condition} is satisfied, two BHs can become bounded and form a BBH. They orbit around each other with an almost periodic elliptical motion, emitting periodic GWs which carry energy out of the system. This leads to a progressive decrease of the distance between BHs and, eventually, to their merger. However, it has been shown that, in the dense environment, a third BH interacts with the BBH within its evolution, breaking the binary system and avoiding the merger~\cite{Eroshenko:2016hmn,Raidal:2018bbj,Vaskonen:2019jpv}. Thus, the possible scenario of BBH formation is that they are formed at the cluster center, and those who are ejected to the outskirt of the cluster eventually merge and emit GWs. 

The cross-section for BH encounters is given by the impact parameter $b$ and can be written as $\sigma = \pi b^2 = \pi (GM/v_0^2)^2(e^2-1)$. By substituting $(e^2-1)$ which satisfies the condition Eq.~\eqref{eq:condition} with the approximation of $e\approx 1$, we obtain the cross-section for forming a BBH
\ba
\sigma^{\rm BBH} = \pi\left(\frac{340\pi}{3}\right)^{2/7}
\frac{G^2M^{10/7}(m_1\,m_2)^{2/7}}{c^{10/7}v_0^{18/7}}\,.
\ea
This gives the rate of forming one BBH as $\tau_{\rm ind}= n(m) v_{\rm PBH} \sigma^{\rm BBH}$, where $n(m)$ is the number density of PBHs and $v_{\rm PBH}=v_0/\sqrt{2}$ is the PBH velocity. For PBHs clustered in dense halos, using the local density contrast $\delta_{\rm loc}$, we can write the number density as $n(m)\equiv\delta_{\rm loc}\overline{\rho}_{\rm DM}/m$, with the mean DM cosmological energy density $\overline{\rho}_{\rm DM}=\Omega_{\rm DM}\rho_c$ where $\Omega_{\rm DM}\simeq 0.25$ is the density parameter for DM. 

The total merger rate per comoving volume can be obtained by multiplying the number of PBHs in the comoving volume. Note that we multiply the averaged number density $n(m)/\dloc$ rather than the clustered one $n(m)$ in order to obtain merger rate in the Gpc volume. By taking into account a mass distribution of BHs, the total merger rate is given by~\cite{Clesse:2016ajp}
\ba
&& \frac{d\tau^{\rm BBH}}{d\ln m_1\,d\ln m_2} 
= \frac{1}{\delta_{\rm loc}}\,\sigma^{\rm BBH}\,v_{\rm PBH}\,n(m_1)\,n(m_2)
\nn \\
& \approx & 14.8~{\rm yr}^{-1}{\rm Gpc}^{-3} 
h_{70}^4 
\left(\frac{\Omega_{\rm DM}}{0.25}\right)^2
\left(\frac{\delta_{\rm loc}}{10^8}\right) \\
&&\left(\frac{v_0}{10~{\rm km/s}}\right)^{-11/7} f(m_1)\,f(m_2)\frac{M^{10/7}}{(m_1\,m_2)^{5/7}}\,, \nn
\ea
where $f(m)$ is the logarithmic mass function of PBH, such that $\int d m/m \,f(m) = \fPBH \leq 1$. Here, $\fPBH$ is the fraction of DM made of PBH, and we assume $\fPBH = 1$ for simplicity. Typical values for $v_0$ are a few tens of km/s and $\delta_{\rm loc}$ can be taken of order $10^8$, as in Ref.~\cite{Clesse:2016ajp}.

For the energy emission, we follow the formalism of~\cite{Ajith:2009bn}, which includes the contributions of the inspiral, merger, and ringdown parts of the BBH waveform. The energy released to GWs (for $e=0$ and no spin) is given by
\be
\frac{dE^{\rm BBH}}{d\ln f_r} =
\frac{ (\pi G)^{2/3} m_1\,m_2}{3 c^2 M^{1/3}} f_r^{2/3} {\cal F}(f_r)\,,
\ee
where ${\cal F}(f_r)$ is the function to describe the deviation from the frequency dependence of the inspiral phase $f_r^{2/3}$ (see~\cite{Ajith:2009bn,Braglia:2021wwa} for the detailed functional form).

Assuming that the merger rate is constant in time, the redshift integration of Eq.~\eqref{eq:OGW} can be reduced to a simple form. For $\OM=0.31$ (density parameter of matter), we find
\be
\int_0^\infty \frac{dz}{(1+z)^{4/3}H(z)} =
0.76\,H_0^{-1}\,.
\ee

Then, setting ${\cal F}(f_r)=1$, the energy spectrum for the low-frequency inspiral regime can be estimated as
\ba
\label{eq:OGWBBH}
&&\OGW^{\rm BBH}(f) 
\approx
2.39\times 10^{-13}\,
h_{70} \nn \\
&&
\times\left(\frac{\Omega_{\rm DM}}{0.25}\right)^2
\left(\frac{\delta_{\rm loc}}{10^8}\right)
\left(\frac{v_0}{10~{\rm km/s}}\right)^{-11/7}
\left(\frac{f}{{\rm Hz}}\right)^{2/3} \\
&&
\times\int\!dm_1\,dm_2 
\frac{f(m_1)\,f(m_2)\,
(m_1+m_2)^{23/21}}
{(m_1\,m_2)^{5/7}}
\nn
\,,
\ea
with $m_i$ in solar masses.

We can refine the analysis by assuming a redshift dependence of the merger rate, $\tau^{BBH}\propto(1+z)^\beta$, with exponent $0<\beta<1.28$~\cite{Mukherjee:2021ags,Raidal:2018bbj}. In this case, it can be shown that the low frequency part of the spectrum gets enhanced by at most a constant factor 3.9, slightly modifying the shape of the spectrum near the peak. However, the slope at low frequencies stays at $f^{2/3}$ and the cutoff at high frequencies remains invariant.

\subsection{II.2 Close hyperbolic encounters}
In the dense environment at the center of the cluster, a large fraction of BH encounters does not end up producing bounded systems, but rather produces a single scattering event. These encounters have got less attention than BBHs, but in fact, CHE should be more common at the inner part of BH clusters~\cite{Trashorras:2020mwn}, and they emit GWs, which should be considered both for individual events~\cite{Garcia-Bellido:2017knh, Garcia-Bellido:2017qal} and for their contribution to the SGWB. 

If a pair of BHs do not satisfy the condition Eq.~\eqref{eq:condition}, then the two BHs pass away, and it becomes a CHE event. The cross-section is again given by $\sigma = \pi b^2 = \pi (GM/v_0^2)^2(e^2-1)$, but this time, we do not have the condition imposed on the eccentricity coming from Eq.~\eqref{eq:condition}. Thus, apart from the BH masses, we have two parameters $v_0$ and $e$ for CHE. Then the total event rate is given by 
\ba
&& \frac{d\tau^{\rm CHE}}{d m_1\,d m_2}  
= \frac{1}{\dloc} \sigma v_{\rm PBH}  n(m_1) n(m_2) \nn \\
& \approx & 25.4 \times 10^{-8}~{\rm yr}^{-1}{\rm Gpc}^{-3}\,h_{70}^4
\left(\frac{\ODM}{0.25}\right)^2
\left(\frac{\dloc}{10^8}\right)
 \nn \\
&& \times \frac{f(m_1)}{m_1}\frac{f(m_2)}{m_2}\frac{M^2}{m_1\,m_2}
\frac{e^2-1}{(v_0/c)^3}\,.
\label{Eq:CHErate}
\ea
Note that here again we have multiplied the averaged number density to obtain the total rate (see Appendix for further discussion). The relative velocity $v_0$ can be related to the semi-major axis as $a=G M/v_0^2$, which from now on will be used instead of $v_0$. We will also include the factor $(1+z)^\beta$ in order to parametrize a possible time dependence of the event rate.

In the case of CHE, the energy emitted per logarithmic frequency bin is given by~\cite{DeVittori:2012da,Garcia-Bellido:2017qal}
\be
\frac{dE^{\rm CHE}_{\rm GW}}{d\ln f_r} = \nu\frac{dE_{\rm GW}}{d\nu} =
\frac{4\pi}{45}\,\frac{G^3m_1^2m_2^2}{a^2c^5\nu_0}\,\nu^5F_e(\nu)\,,\\[1mm]
\ee
where we have defined $\nu \equiv 2\pi\nu_0\,f_r$ and $\nu_0^2 \equiv a^3/GM$. The frequency dependence can be approximated by~\cite{Garcia-Bellido:2017knh} 
\ba
\nu^5F_e(\nu) &\simeq& \frac{12F(\nu)}{\pi\,y\,(1+y^2)^2}\,e^{-2\nu\,\xi(y)} \,, \nn \\
F(\nu) &=& \nu^2\left(1-y^2-3\,\nu\,y^3+4\,y^4
+ 9\,\nu\,y^5+6\,\nu^2 y^6\right) \,, \nn \\
\xi(y) &=& y - {\rm tan}^{-1}y \,, \nn\\
y &=& \sqrt{e^2-1} \,.
\ea
In order to perform the redshift integration of Eq.~\eqref{eq:OGW}, we define the following function 
\ba
I[y,\,x_0] &\equiv&\frac{\pi}{12}
\int_0^\infty dz\,\frac{\nu^5\,F_e(\nu)\,H_0\sqrt{\OM}}{(1+z)^2H(z)}(1+z)^\beta \nn \\
&\simeq&
\frac{\pi x_0^{5/2-\beta}}{12}\int_{x_0}^\infty d\nu\,\nu^{3/2+\beta}F_e(\nu) \nn \\[2mm]
&=&\frac{2x_0^{5/2-\beta}}{(2\xi)^{3/2+\beta}}\frac{1}{y(1+y^2)^2}\times \nn \\
&& \left[ 2(1-y^2+4y^4)\,\xi^2\,\Gamma\left(-\frac{1}{2}+\beta,\,2x_0\xi\right)\right. \nn \\
&& +3y^3(-1+3y^2)\,\xi\,\Gamma\left(\frac{1}{2}+\beta,\,2x_0\xi\right) \nn \\
&& \left.+3y^6\Gamma\left(\frac{3}{2}+\beta,\,2x_0\xi\right)\right],
\label{eq:I_CHE}
\ea
where we have defined $x_0\equiv 2\pi \nu_0 f$ so that $\nu=x_0(1+z)$ and $\Gamma$ is the upper incomplete gamma function. We find that the function follows $x_0^\alpha\propto f^\alpha$ at low frequencies, where $\alpha=\min\{2,5/2-\beta\}$. This is one of the key features of the CHE contribution to the SGWB: a measurement of the slope of the low-frequency tail would provide information about the redshift dependence of the event rate, unlike for BBH, which only shifts the amplitude without changing the slope. Since astrophysical and primordial black holes have very different event rates as a function of redshift, this could help distinguish between them.

For the subsequent discussion, we will focus on the constant event rate case $\beta=0$. This allows us to simplify the previous integral as
\be
I[y,\,x_0]\simeq 2\,x_0^2\,e^{-2x_0\xi(y)}
\left(\frac{1-y^2+4y^4+\frac{3}{2}\frac{x_0y^6}{\xi(y)}}{y\,(1+y^2)^2}\right)\,. 
\ee
We can see that this function peaks at around $x_0^{\rm peak} \simeq 1/\xi(y)$ and decays as $\exp(-2x_0\xi(y))$ at higher frequencies. Using the approximation of $\xi(y)\approx y^3/3$ for $y\ll 1$, the peak frequency can be estimated as
\be
\label{eq:fmax}
f_{\rm peak} \simeq 43\,{\rm Hz}\,\left(\frac{y}{0.01}\right)^{-3}
\left(\frac{M}{200\Msun}\right)^{1/2}
\left(\frac{a}{0.1~{\rm AU}}\right)^{-3/2}\,,
\ee
and thus depends on the intrinsic properties of the CHE ($M,\,a,\,e$).
Putting all of this together and redefining $I=2\,x_0^2\,y^{-1}\,\exp[-2 x_0\xi(y)]\,\tilde{I}$, so that $\tilde{I}$ is of order unity at low frequencies, we find
\ba
\label{eq:OGWCHE}
\OGW^{\rm CHE}(f) &\approx&
9.81\times 10^{-13}\,
h_{70}\left(\frac{\Omega_{\rm M}}{0.3}\right)^{-1/2}
\left(\frac{\Omega_{\rm DM}}{0.25}\right)^2 \nn \\
&& \hspace{-2cm}\times
\left(\frac{\delta_{\rm loc}}{10^8}\right)
\left(\frac{a}{0.1~{\rm AU}}\right)\left(\frac{f}{10\rm Hz}\right)^2 \left(\frac{y}{0.01}\right) \nn \\
&& \hspace{-2cm}
\times\int\!\frac{dm_1}{100\Msun}\,\frac{dm_2}{100\Msun} f(m_1)\,f(m_2)\,
\,e^{-2x_0\xi(y)}\,\tilde{I}[y,x_0]\,.
\ea

For the peak frequency~\eqref{eq:fmax}, we can approximate
\ba
\label{eq:OGW_CHE_max}
\OGW^{\rm CHE}(f_{\rm peak}) &\approx&
3.6\times 10^{-13}\,
h_{70} \nn \\
&& \hspace{-2cm}\times
\left(\frac{\Omega_{\rm M}}{0.3}\right)^{-1/2}
\left(\frac{\Omega_{\rm DM}}{0.25}\right)^2
\left(\frac{\delta_{\rm loc}}{10^8}\right)
\left(\frac{a}{0.1~{\rm AU}}\right)^{-2} \nn \\
&& \hspace{-2cm}
\times \left(\frac{y}{0.01}\right)^{-5}\, \frac{m_1}{100\Msun}\,\frac{m_2}{100\Msun}\,\frac{m_1+m_2}{200\Msun}\,.
\ea
For instance, by taking $m_1=m_2=300\Msun$, $a=5$AU and $y=2\times 10^{-3}$, we find $\OGW(f_{\rm peak})\approx 1.2\times 10^{-11}$ at the LIGO frequency band, $f_{\rm peak}\approx 26$ Hz. On the other hand, $m_1=m_2=300\Msun$, $a=5\times 10^7$AU and $y=10^{-5}$ yields $\OGW(f_{\rm peak})\approx 3.9\times 10^{-14}$ at the LISA frequency band, $f_{\rm peak}\approx 6.7$ mHz.

Eq.~\eqref{eq:OGW_CHE_max} indicates the intuitively expected behavior that we get larger SGWB amplitude for larger mass, smaller semi major axis, and smaller $y$ (eccentricity close to unity). These parameter values would be distributed on a wide range and in principle we should marginalize them over for obtaining the SGWB spectrum. Here, for simplicity, we assume a log-normal distribution of median $m_0/a_0/y_0$ and its variance $\sigma_m/\sigma_a/\sigma_y$.

\subsection{III. Comparison between both cases}

\begin{figure}[t]
\includegraphics[width=0.95\linewidth]{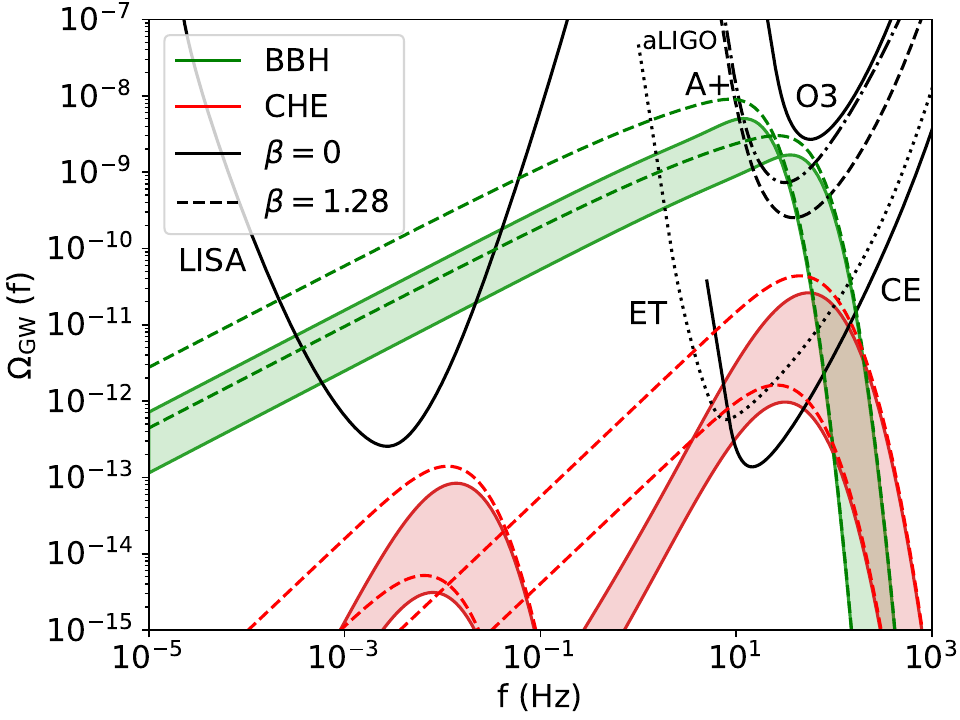}
\caption{\label{fig:OGW} Comparison of the SGWB spectrum originating from BBHs and CHEs, both for $\beta=0$ (solid) and 1.28 (dashed). We also plot the power-law integrated sensitivity curves of several GW detectors for a signal-to-noise ratio of $10$ and an observation time of $1$ year, following the formalism in~\cite{Thrane:2013oya}. For the BBH curves, we take $m_1=m_2=100-300~\Msun$ and $v_0=30$km/s. The CHE curves correspond to the same range of masses with $a_0=5$AU, $y_0=2\times 10^{-3}$ for $\sim 10$Hz, and $a_0=5\cdot 10^7$AU, $y_0=10^{-5}$ for the mHz range. For all cases, we take $\sigma_a,\sigma_y = 0.1$, $\sigma_m = 0.5$, and $f_{\rm PBH}=1$. For a smaller fraction of PBHs, the GW spectral amplitude simply scales as $\Omega_{\rm GW}\propto f^2_{\rm PBH}$.}
\end{figure}

In order to understand the differences between the BBH and CHE cases, in Fig.~\ref{fig:OGW}, we plot the GW spectra. We have marginalized over mass distributions centered at $100-300\Msun$, in order to show reasonable contributions in both the LISA and LIGO frequency bands. However, one could in principle consider other mass ranges or distributions. As illustrated by Eqs.~\eqref{eq:fmax} and~\eqref{eq:OGW_CHE_max}, higher masses will produce greater backgrounds peaking at higher frequencies, while the opposite will be true for lower masses.

The first thing we notice in Fig.~\ref{fig:OGW} is that the SGWB spectrum from CHEs is steeper than the one from BBHs, which could be deduced from the frequency dependence of Eqs.~\eqref{eq:OGWBBH} and~\eqref{eq:I_CHE}. In the case of the constant merger rate ($\beta=0$), the SGWB spectrum inherits the $f^{2}$ dependence of the spectrum of the individual event at low frequencies, while the individual BBH spectrum has the $f^{2/3}$ dependence.
One possible physical interpretation is that a CHE emits a single burst of GWs, whereas the ones from a BBH are essentially periodic with increasing frequency. It is, therefore, natural that the BBH case has a broader spectrum in frequency. 

In addition, we can see the different behavior of the BBH and CHE tails when merger rate increases towards high redshift. As was mentioned before, the change of the BBH curves with $\beta$ is just an overall enhancement up to a factor 3.9, whereas the CHE curves notably modify their slopes. This sensitivity of the CHE background to $\beta$ is very relevant, since the rate dependence on redshift is one of the features that distinguish astrophysical from primordial black holes~\cite{Ajith:2009bn}. Therefore, a measurement of the tail of the CHE background could provide useful information about the relative abundance of both populations.

For the LIGO frequency range, with modest values of the parameters, the CHE spectrum can reach ET and CE sensitivities, as can be seen in~Fig.\ref{fig:OGW}. Although the CHE contribution is below the BBH curves, note that the destruction of the binary system in a dense environment, which could lower the GW amplitude, is not taken into account. Note also that some BBH events with a large signal can be detected individually and subtracted from the data~\cite{Regimbau:2016ike}, which may allow us to probe the other SGWB component below the curve. This may also be the case for the more modest CHE curves in the LISA range even if they fall below its sensitivity curve, given the expected improved sensitivity of the LISA-TianQin/Taiji networks~\cite{Liang:2021bde, Wang:2021uih,Wang:2021njt}.

More restrictive parameters are needed in order to produce a relevant CHE background in the LISA frequency band. This difficulty can easily be understood if we rewrite Eq.~\eqref{eq:OGW_CHE_max} in terms of the peak frequency,
\ba
\OGW^{\rm CHE}(f_{\rm peak}) & \approx & 4.4\times 10^{-13}\,
h_{70} \nn \\
&& \hspace{-2cm}\times\left(\frac{\Omega_{\rm M}}{0.3}\right)^{-1/2}
\left(\frac{\Omega_{\rm DM}}{0.25}\right)^2
\left(\frac{\delta_{\rm loc}}{10^8}\right)
\left(\frac{f_{\rm peak}}{50~{\rm Hz}}\right)^{4/3} \nn \\
&& \hspace{-2cm}\times \left(\frac{y}{0.01}\right)^{-1}\, \frac{m_1}{100\Msun}\,\frac{m_2}{100\Msun}\,\left(\frac{m_1+m_2}{200\Msun}\right)^{1/3}.
\ea

We can thus see that, for fixed masses and eccentricity, the maximum amplitude $\OGW^{\rm CHE}(f_{\rm peak})$ grows with $f_{\rm peak}^{4/3}$. Therefore, with modest parameters as the CHE curves in the LIGO range, the amplitude of the background decays significantly if we try to translate the curve to the LISA range. On the other hand, however, this is a hint that the SGWB from CHE may play an important role in higher frequency ranges, such as the ultra-high frequency (UHF, MHz-GHz) band for which there are good prospects of detection in the future~\cite{Aggarwal:2020olq}.

\section{IV. Conclusions}
In this paper, we have proposed a new source of SGWB from CHEs. We have computed the SGWB spectrum from a superposition of GWs from CHE events and compared the amplitude with the one from BBHs. We have seen that they have different frequency dependencies, which would help to distinguish the two different origins when detection of SGWB is made. Furthermore, as shown in Fig.~\ref{fig:OGW}, we have found that there exist combinations of parameter values that can make the CHE contribution detectable by future GW interferometers, especially with ET, CE or UHF experiments and with more difficulty in the LISA range.

In addition, we discovered that a change on the event rate dependence on redshift translates into a change of slope for the CHE contribution of the low-frequency tail. This is something that doesn't happen for BBH and opens the possibility to probing the time evolution of the event rate.

As we have discussed, formations of BBHs and CHEs strongly depend on the clustering nature of the PBHs. Current cosmological observations have not yet provided a clear picture of BH distribution in the Universe. Detection of SGWB would provide new implications on the BH evolution. Note that, although we have focused on SGWB from PBHs in this paper, GWs from astrophysical BHs (ABHs) also form a SGWB, and we would observe the incoherent sum of all SGWB. Given the fact that the event rate evolves differently with time for ABHs and PBHs, the slope of the tail of the CHE background could serve to disentangle both contributions and derive their relative abundance. The combination of other information from further investigation, such as spectral shape~\cite{Kuroyanagi:2018csn,Caprini:2019pxz,Mukherjee:2021ags}, anisotropy~\cite{Cusin:2019jhg,Contaldi:2020rht,KAGRA:2021mth}, and popcorn feature~\cite{Mukherjee:2019oma,Smith:2020lkj}, will also help to obtain implications on the origin of BHs. 

As a final remark, although we have made an estimation using the simplified picture, where we assume uniform over density region with the parametrization of $\dloc$, our work can be extended to incorporate a more detailed clustering profile of PBHs. Besides, we could consider more realistic distributions of the CHE parameters, such as semi-major axis and eccentricity. In fact, we have observed that when we make the log-normal distribution of the parameters wider, the spectral shape changes dramatically, and the peak amplitude tends to get enhanced. We leave the detailed analysis for future work.

\section*{Acknowledgements} 

The authors acknowledge support from the Research Project PGC2018-094773-B-C32 [MINECO-FEDER], and the Centro de Excelencia Severo Ochoa Program SEV-2016-0597. S.J. is supported by the FPI grant PRE2019-088741 funded by MCIN/AEI/10.13039/501100011033. S.K. is supported by the Atracción de Talento contract no. 2019-T1/TIC-13177 granted by the Comunidad de Madrid and Japan Society for the Promotion of Science (JSPS) KAKENHI Grant no. 20H01899 and 20H05853. We would like to thank the referee for useful comments on the manuscript.

\appendix
\section*{Appendix: Clumping factor} 
The clumpiness of PBH halos is an important factor for enhancing the merger rate of CHE, and accordingly, the GW amplitude. We see in Eq.~\eqref{Eq:CHErate} that the enhancement is by $\delta_{\rm loc}$, instead of $\delta_{\rm loc}^2$ indicated by~\cite{Garcia-Bellido:2017knh, Garcia-Bellido:2017qal}. The difference arises because we multiply the individual merger rate by the averaged PBH density instead of the local density for obtaining the total merger rate.

Let us derive the dependence of $\propto\delta_{\rm loc}$ in another way. 
In our setting, the density contrast in a halo is uniformly described by $\delta_{\rm loc}$, while some articles~\cite{OLeary:2008myb,Mukherjee:2020hnm} estimate the event rate by taking into account the halo profile and halo number density. The analogy can be seen in the gamma-ray background from DM annihilation. In~\cite{Ahn:2004yd}, the clumping factor is defined as $C_X(z)\equiv\langle\rho_X^2\rangle_z/\langle\rho_X\rangle_z^2$, and given by
\begin{equation}
C_X(z)=\frac{(1+z)^3}{\langle\rho_X\rangle_z^2}\int^\infty_{M_{H{\rm min}}} dM_H\frac{dn(M_H,z)}{dM_H}\int d^3 r \rho_X^2(M,r) \\,
\label{Eq:4}
\end{equation}
where $\rho_X(M_H,r)$ is the DM density profile, $\langle\rho_X\rangle$ is the averaged DM density in the Universe, and $d n(M_H,z)/dM_H$ is the comoving number density of halos in the mass range of $M_H \sim M_H + dM_H$.  

Let us apply this formula in our setting where the DM density is constant within a halo and the density profile is described as $\rho_X(M_H,r)=\delta_{\rm loc}\langle\rho_X\rangle_z \equiv\rho_H  ~ ({\rm for~} r< r_H)$ with $r_H$ being the halo size. Assuming that all the halos have the same mass $M_H$, the volume of a halo is given by $V_H=M_H/\rho_H=M_H/(\delta_{\rm loc}\langle\rho_X\rangle_z)$, and the number density of halo is $n(M_H,z)=\langle\rho_X\rangle_z/M_H\equiv n_{M_H}$. Applying them to \eqref{Eq:4}, we can find the clumping factor is $\delta_{\rm loc}$ as the following,
\begin{eqnarray}
C_X (z=0) &\sim& \frac{1}{\langle\rho_X\rangle_z^2} n_{M_H} V_H \rho_H^2 \nn \\
&=& \frac{1}{\langle\rho_X\rangle_z^2}\frac{\langle\rho_X\rangle_z}{M_H}\frac{M_H}{\delta_{\rm loc}\langle\rho_X\rangle_z}(\delta_{\rm loc}\langle\rho_X\rangle_z)^2 \nn \\ 
&=& \delta_{\rm loc}
\end{eqnarray}
 
\bibliography{SGWB}

\end{document}